\documentclass[12pt]{iopart}

\usepackage{epsfig}
\usepackage{multirow}
\usepackage{iopams}

\begin{document}

\vspace*{-2cm}

\centerline{\large {\bf A simplified model of the source channel}}
\centerline{\large {\bf of the Leksell Gamma Knife$^{\circledR}$: testing}}
\centerline{\large {\bf multisource configurations with PENELOPE}}

\vspace{.3cm}

\begin{center}
{\bf Feras M.O. Al-Dweri and Antonio M. Lallena}\\
Departamento de F\'{\i}sica Moderna, 
Universidad de Granada, E-18071 Granada, Spain.
\end{center}

\vspace{.2cm}

\small{A simplification of the source channel geometry of the Leksell Gamma
Knife$^{\circledR}$, recently proposed by the authors and checked for
a single source configuration (Al-Dweri \etal 2004), has been used to
calculate the dose distributions along the $x$, $y$ and $z$ axes in a
water phantom with a diameter of 160~mm, for different configurations
of the Gamma Knife including 201, 150 and 102 unplugged sources. The
code PENELOPE (v. 2001) has been used to perform the Monte Carlo
simulations. In addition, the output factors for the 14, 8 and 4~mm
helmets have been calculated. The results found for the dose profiles
show a qualitatively good agreement with previous ones obtained with
EGS4 and PENELOPE (v. 2000) codes and with the predictions of
GammaPlan$^{\circledR}$. The output factors obtained with our model
agree within the statistical uncertainties with those calculated with
the same Monte Carlo codes and with those measured with different
techniques. Owing to the accuracy of the results obtained and to the
reduction in the computational time with respect to full geometry
simulations (larger than a factor 15), this simplified model opens the
possibility to use Monte Carlo tools for planning purposes in the
Gamma Knife$^{\circledR}$.}

\vspace{.3cm}

\section{Introduction}

Leksell Gamma Knife$^{\circledR}$ (GK) is a high precision device
designed to perform radiosurgery of brain lesions. It uses the
radiation generated by 201 $^{60}$Co sources in such a way that a high
dose can be deposited into the target volume with an accuracy better
than 0.3~mm (Elekta 1992), while the critical brain structures
surrounding the lesion to be treated can be maintained under low dose
rates.  The four treatment helmets available together with the
possibility to block individual sources permit to establish optimal
dose distributions. These are determined by means of the
GammaPlan$^{\circledR}$ (GP), the computer-based treatment planning
system accompanying GK units (Elekta 1996).

To ensure GP quality, and due to the usual difficulties in
measuring the physical doses, Monte Carlo (MC) calculations have
played a relevant role as a complementary tool. Most of the
simulations performed (Cheung \etal 1998, 1999a, 1999b, 2000, Xiaowei
and Chunxiang 1999, Moskvin \etal 2002) have pointed out a good
agreement with the GP predictions in the case of a homogeneous phantom.
However, differences reaching 25\% are found for phantoms with
inhomogeneities, near tissue interfaces and dose edges (Cheung \etal
2001).

In a recent paper (Al-Dweri \etal 2004), we have proposed a simplified
model of the source channel of the GK. This model is based on the
characteristics shown by the beams after they pass trough the
treatment helmets. In particular, the photons trajectories, reaching
the output helmet collimators at a given point $(x,y,z)$, show strong
correlations between $\rho=(x^2+y^2)^{1/2}$ and their polar angle
$\theta$, on one side, and between $\tan^{-1}(y/x)$ and their
azimuthal angle $\phi$, on the other. This permits to substitute the full
source channel by a point source, situated at the center of the active
core of the GK source, which emits photons inside the cone defined by
itself and the output helmet collimators. This simplified model
produces doses in agreement with those found if the full geometry of
the source channel is considered (Al-Dweri \etal 2004).

In this work we want to complete the test of this simplified model by
calculating the dose distributions in a water phantom which is
irradiated by a GK unit with different multisource configurations. The
version 2001 of the code PENELOPE (Salvat \etal 2001) has been used to
perform the Monte Carlo simulations. In addition to the dose profiles
around the isocenter of the GK, the output factors for the 4, 8 and
14~mm helmets have been obtained. Our findings have been compared with
various results obtained with other MC codes, such as EGS4 (Cheung
\etal 1999a, 1999b, 2000, Xiaowei and Chunxiang 1999) or PENELOPE
(Moskvin \etal 2002), predicted with GP (Elekta 1992, 1998) or
measured with different techniques (Ma \etal 2000, Tsai \etal 2003).

\section{Material and Methods}

\subsection{Leksell Gamma Knife$^{\circledR}$ model}

\begin{figure}
\begin{center}
\epsfig{figure=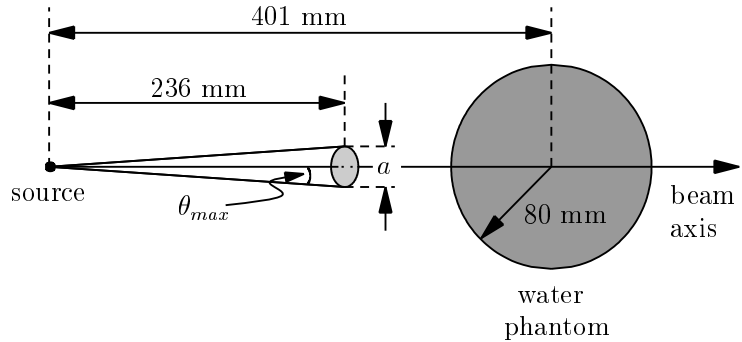,width=10cm}
\end{center}
\caption{Schematic view of the simplified model of the source channel 
of the GK. A point source emits initial photons in a cone defined by
itself and the apertures $a$, which are given in
table \protect\ref{tab:helmets}. The water phantom used to simulate 
the patient head is also drawn. 
\label{fig:simple}}
\end{figure}

Figure \ref{fig:simple} shows a scheme of the simplified model of the
source channel of the GK. It consists in a point source emitting the
initial photons in the cone defined by itself and the helmet outer
collimators, whose apertures $a$ are given in table \ref{tab:helmets},
together with the maximum polar angle, $\theta_{\rm max}$,
corresponding to each helmet. The water phantom is a sphere with 80~mm
of radius. It simulates the patient head and its center coincides with the
isocenter of the GK.

\Table{Apertures, $a$, and maximum polar angles, $\theta_{\rm max}$, of
the emission cones for the initial photons in the simplified model of
the source channel of the Leksell Gamma Knife$^{\circledR}$.  The
values of the apertures correspond those of the helmet outer
collimators and have been taken from Moskvin \etal (2002).
\label{tab:helmets}}
\br
Final beam diameter & 4~mm & 8~mm & 14~mm & 18~mm \\
\mr
$a$ [mm]  & 2.5 & 5.0 & 8.5 & 10.6 \\
$\theta_{\rm max}$ [deg] & 0.303 & 0.607 & 1.032 & 1.287 \\
\br
\end{tabular}
\end{indented}
\end{table}

\begin{figure}
\begin{center}
\epsfig{figure=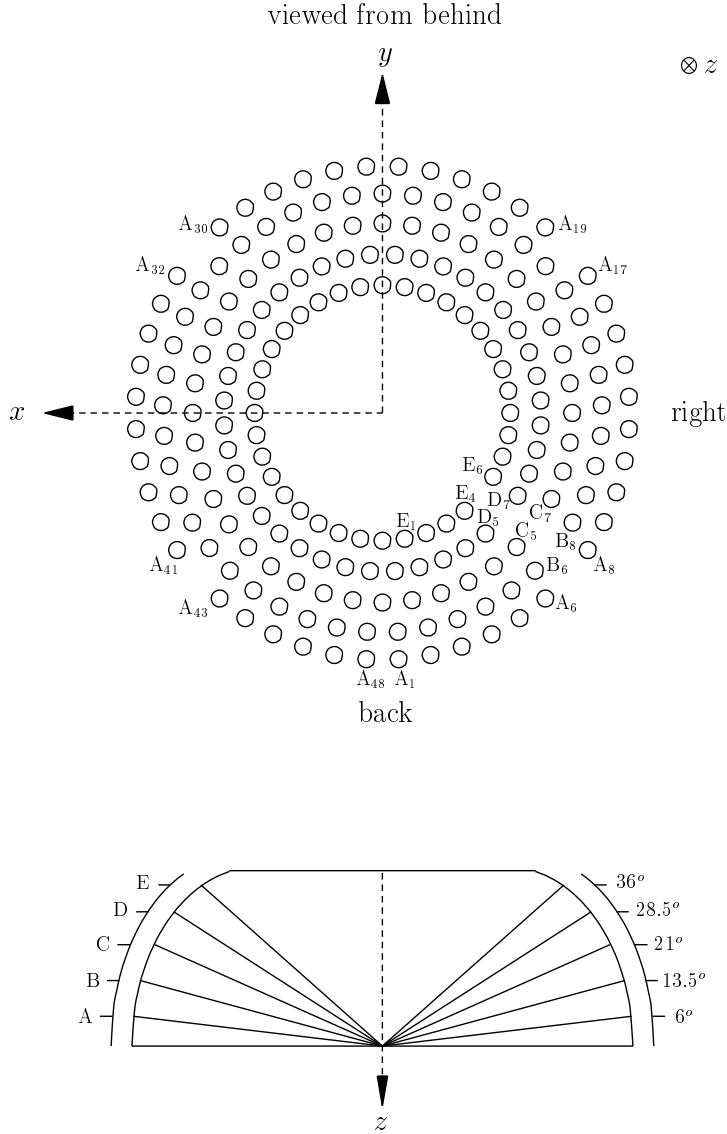,width=10cm}
\end{center}
\caption{Scheme of the situation of the 201 sources of the GK. The
reference system used is indicated in the upper panel. Also the
labeling of the sources is included. In the lower panel the
elevation angles of the five rings in which the sources are
distributed are shown.
\label{fig:GK}}
\end{figure}

This simplified geometry has been considered for all the 201 source
channels of the GK. Figure \ref{fig:GK} shows a scheme of the
situation of these sources. In the upper panel, the reference system
we have considered is indicated. The origin of coordinates is situated
at the isocenter of the GK and the $z$ axis is in the patient axis
pointing from the head to the feet. The lower panel shows the disposal
of the five rings in which the sources are distributed as well as the
elevation angles of each one with respect to the isocenter
plane. There are 44 sources in rings A and B, 39 in rings C and D and
35 in ring E. Table \ref{tab:source-coor} shows the spherical
coordinates of the 201 point sources. All of them have $r=401$~mm. On
the other hand, the angles $\phi$ for each source in a given ring
$\alpha$ are given by $\phi_i^\alpha=\phi_1^\alpha-i\Delta
\phi^\alpha$, $i$ being the order label shown in the upper panel of
figure \ref{fig:GK}. 

\Table{Spherical coordinates of the 201 point sources assumed in our
calculations. We give the $\theta^\alpha$ coordinates of the five
rings A to E in which the sources are distributed. The coordinates
$\phi$ of each source are obtained as 
$\phi_i^\alpha=\phi_1^\alpha-i\Delta \phi^\alpha$, $i$ being the
order label as shown in Fig. 2. All the sources have $r=40.1$~cm.
\label{tab:source-coor}}
\br 
             $\alpha$ & {A} & {B} & {C} & {D} & {E} \\ \br 
$\theta^\alpha$~[deg] & 96.0 & 103.5 & 111.0 & 118.5 & 126.0 \\ \mr 
$\phi^\alpha_1$~[deg] & 266.25 & 266.0 & 261.0 & 255.5 & 260.0 \\ \mr 
$\Delta \phi^\alpha$~[deg] & 7.5 & 8.0 & 9.0 & 9.0 & 10.0 \\  
\br
\end{tabular}
\end{indented}
\end{table}

It should be noted that the distribution of the sources in the rings
is not completely uniform because some of them are not present (for
example those labeled A$_7$, B$_7$, C$_6$, D$_6$, E$_5$, etc.) and
this breaks the cylindrical symmetry of the system. Thus, the doses we
have calculated depend on the three cartesian coordinates,
$D(x,y,z)$. The scoring voxels we have used present $\Delta x=$0.5~mm
and $\Delta y=\Delta z=1$~mm, for the 18 and 14~mm helmets, and
$\Delta x=$0.25~mm and $\Delta y=\Delta z=$0.5~mm, for the 8 and 4~mm
ones.

In connection with this point, we have studied the asymmetry between
$x$ and $y$ axes, by calculating the quantity
\begin{equation}
A_{xy} (s) \, = \, \displaystyle
\frac{D(s,0,0)-D(0,s,0)}{D(s,0,0)} \, ,
\label{eq:asymm}
\end{equation}
and, also, the asymmetries between the
negative and positive parts of the $x$ axis,
\begin{equation}
A_{x} (s) \, = \, \displaystyle
\frac{D(s,0,0)-D(-s,0,0)}{D(s,0,0)} \, , s \geq 0 \, ,
\label{eq:asymm-x}
\end{equation}
and of the $y$ axis,
\begin{equation}
A_{y} (s) \, = \, \displaystyle
\frac{D(0,s,0)-D(0,-s,0)}{D(0,s,0)} \, , s \geq 0 \, .
\label{eq:asymm-y}
\end{equation}

\begin{figure}
\begin{center}
\epsfig{figure=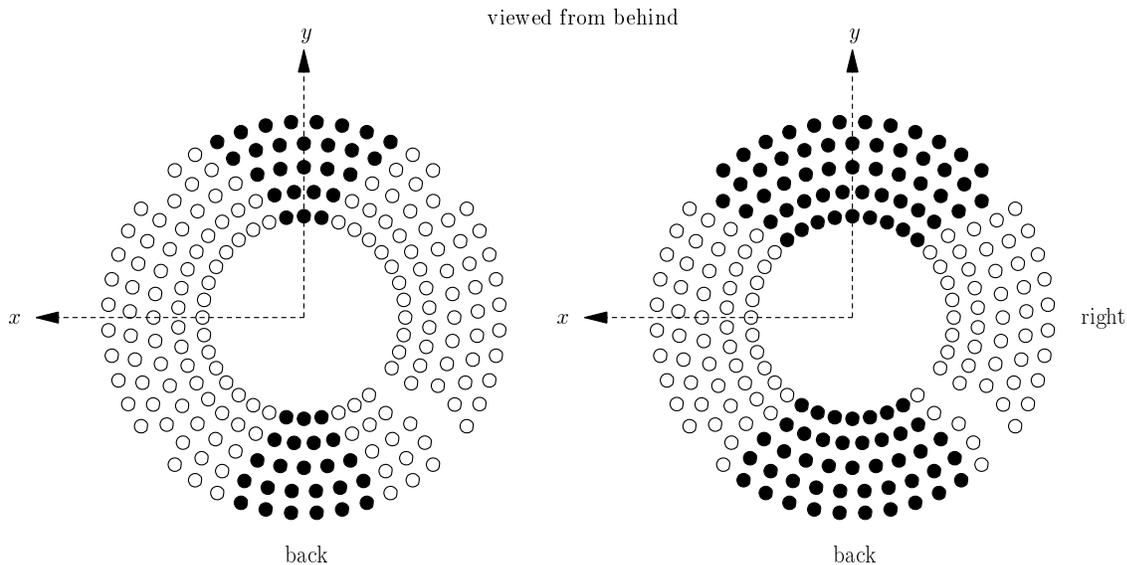,width=15cm}
\end{center}
\caption{Plug patterns for the configurations with 150 (left panel)
and 102 (right panel) unplugged sources considered in this work. The black
circles represents the plugged sources.
\label{fig:plugged}}
\end{figure}

In these calculations we have considered the configuration in which
the 201 sources are unplugged. In addition, configurations with 150 and
102 sources have been considered. Figure \ref{fig:plugged} shows the
corresponding plug patterns taken into account. Therein, the black
circles represent the plugged sources.

We have also calculated the dose output factors for the configuration
with 201 sources. These are defined as the dose rate for a given
collimator helmet, relative to that of the 18~mm helmet, at the
isocenter, in presence of the phantom. To accumulate the energy, we
have considered two kind of voxel centered at the isocenter. First, we
have assumed a cubic voxel with dimensions $(2\Delta x,2\Delta
y,2\Delta z)$, with $\Delta x$, $\Delta y$ and $\Delta z$ equal to the
values given above for the scoring voxels. This corresponds to take
the eight scoring voxels surrounding the isocenter as a unique
voxel. Second, we have considered three spherical voxels with radii
0.5, 0.75 and 1~mm.

To calculate the output factors, the doses obtained 
have been renormalized to the case of a point source emitting
isotropically,
\begin{equation}
D_{\rm norm}(x,y,z) \, = \, f_{\rm norm} \, D(x,y,z) \, ,
\end{equation}
where the normalization factor is
\begin{equation}
f_{\rm norm} \, = \, 
\displaystyle \frac{1-\cos \theta_{\rm max}}{2} \, ,
\end{equation}
with the corresponding $\theta_{\rm max}$ for each helmet.

\subsection{Monte Carlo calculations}

In this work we have used PENELOPE (v. 2001) (Salvat \etal
2001) to perform the calculations. PENELOPE is a general purpose MC
code which permits the simulation of the coupled electron-photon
transport. The energy range in which it can be applied goes from a few
hundred eV up to 1~GeV, for arbitrary materials. PENELOPE describes in
an accurate way the particle transport near interfaces.

PENELOPE performs analog simulation for photons and uses a mixed scheme
for electrons and positrons. In this case, events are classified as
hard (which are simulated in detail and are characterized by polar
angular deflections or energy losses larger than certain cutoff
values) and soft (which are described in terms of a condensed
simulation based on a multiple scattering theory). Details can be
found in Salvat \etal (2001). The full tracking is controlled by means
of five parameters. $C_1$, $C_2$, $W_{\rm cc}$, $W_{\rm cr}$ and
$s_{\max}$. Besides the absorption energies for the different
particles must be supplied. Table \ref{tab:parameters} shows the
values we have assumed in our simulations for these parameters and for
the two materials (air and water) present in the geometry.

\Table{PENELOPE tracking parameters of the two materials assumed in our 
simulations. $E_{\rm abs}$($\gamma$) and $E_{\rm
abs}$(e$^{-}$,e$^{+}$) stand for the absorption energies corresponding
to photons and electrons and positrons, respectively.
\label{tab:parameters}}
\br
materials & ~~ & Air & Water \\ 
\mr
       $E_{\rm abs}$($\gamma$) [keV] && 1.0  & 1.0  \\
$E_{\rm abs}$(e$^{-}$,e$^{+}$) [keV] && 0.1  & 50.0 \\
                             $C_{1}$ && 0.05 & 0.1  \\
                             $C_{2}$ && 0.05 & 0.05 \\
                  $W_{\rm cc}$ [keV] && 5.0 & 5.0 \\
                  $W_{\rm cr}$ [keV] && 1.0 & 1.0 \\
                    $s_{\max}$ [cm]  && $10^{35}$ & $10^{35}$ \\
\br
\end{tabular}
\end{indented}
\end{table}                               

Initial photons were emitted with the average energy 1.25~MeV. For
each history in the simulation, a source was selected by sampling
uniformly between the unplugged sources in the configuration analyzed.
This determines the coordinates of the initial photon and
the beam axis direction. Then the initial photon
is emitted uniformly in the corresponding emission cone as defined in
the simplified geometry.

The number of histories simulated has been chosen in each case to
maintain the statistical uncertainties under reasonable levels.
The uncertainties given throughout the paper correspond to
1$\sigma$.

\Table{Composition of the materials assumed
in the MC simulations performed in this work. The values correspond to
the weight fraction of each element in the material. Also the
densities are quoted.
\label{tab:materials}}
\br
& \centre{1}{Air} & \centre{1}{Water} \\
\mr
H  &          & 0.111894 \\
C  & 0.000124 &          \\
N  & 0.755267 &          \\
O  & 0.231781 & 0.888106 \\
Ar & 0.012827 &          \\
\mr
density [g cm$^{-3}$] & 0.0012048 &  1.0  \\
\br
\end{tabular}
\end{indented}
\end{table}

The simulation geometry has been described by means of the geometrical
package PENGEOM of PENELOPE. Table \ref{tab:materials} gives the
composition and densities of the two materials (air and water) assumed
in our simulations.

To give an idea of the time needed to perform the simulations
discussed below, we can say that it takes 11.2 minutes of CPU
for each $10^6$ histories in a Origin 3400 of Silicon Graphics with a
CPU R14000A at 600~MHz. In a PC with a CPU AMD Athlon XP 1800+ at
1600~Mhz the time needed is 16.6 minutes.

\section{Results}

In our first calculation, the configuration in which all the 201
sources are unplugged has been considered (see figure \ref{fig:GK}). A
total of $15\cdot 10^7$ histories have been followed.

\begin{figure}
\begin{center}
\epsfig{figure=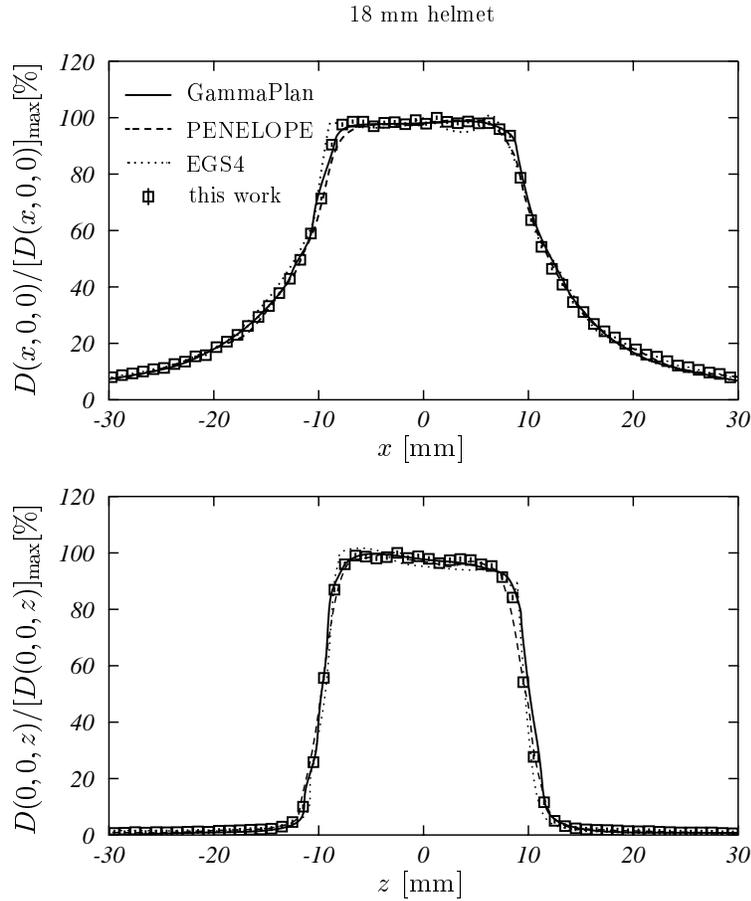,width=10cm}
\end{center}
\caption{Dose profiles at the isocenter, relative to their respective
maxima, in percentage, for the 18~mm helmet. Upper (lower) panel shows
the results along $x$ ($z$) axis. Open squares are the results of our
simulations. Dashed-dotted curves correspond to EGS4 results and have
been obtained by scanning directly figures 4 and 5 of Xiaowei and
Chunxiang (1999). Dashed curves have been obtained by Moskvin \etal
(2002) with PENELOPE for a polystyrene phantom. Solid curve
corresponds to the predictions of GP quoted by Moskvin \etal
(2002). Both results have been obtained from figure 9 of Moskvin \etal
(2002).
\label{fig:201-18}}
\end{figure}

\begin{figure}
\begin{center}
\epsfig{figure=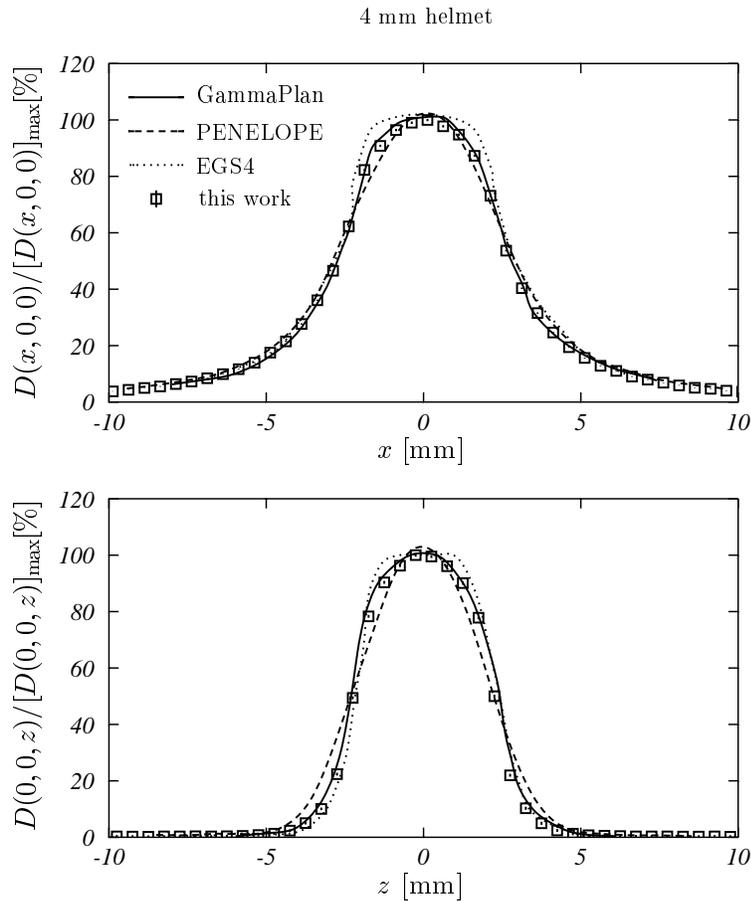,width=10cm}
\end{center}
\caption{The same as in figure \protect\ref{fig:201-18} but for the
4~mm helmet.
\label{fig:201-04}}
\end{figure}

\begin{figure}
\begin{center}
\epsfig{figure=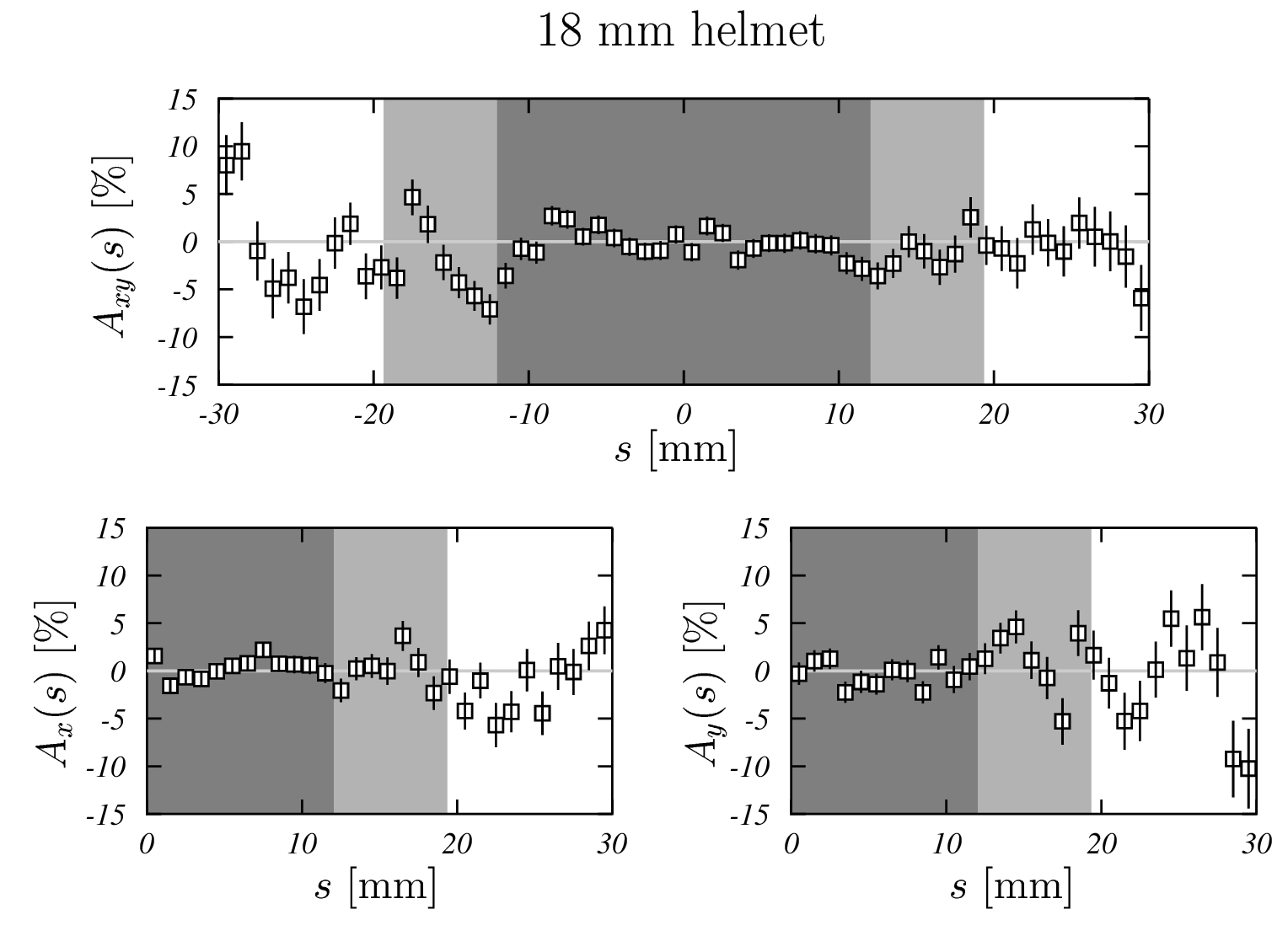,width=12cm}
\end{center}
\caption{Asymmetries $A_{xy}$ (upper panel), $A_{x}$ (lower left
panel) and $A_{y}$ (lower right panel), as given by equations
(\protect\ref{eq:asymm}), (\protect\ref{eq:asymm-x}) and
(\protect\ref{eq:asymm-y}), in percentage, for the 18~mm helmet. The
shadow areas mark the $s$ values for which the dose $D(s,0,0)$ is
larger than 20\% (clearer area) and 50\% (darker area) of the maximum
dose $[D(s,0,0)]_{\rm max}$.
\label{fig:asymm}}
\end{figure}

Figures \ref{fig:201-18} and \ref{fig:201-04} show the dose profiles
at the isocenter, $D(x,0,0)$ (upper panel) and $D(0,0,z)$ (lower
panel), relative to their respective maxima, in percentage, for the 18
and 4~mm helmets, respectively. The results of our simulations
(squares) are compared with those obtained with EGS4 by Xiaowei and
Chunxiang (1999) (dotted curves), with PENELOPE (v. 2000) by Moskvin
\etal (2002) (dashed curves) and with the predictions of GP (Moskvin
\etal 2002) (solid curves).  The values quoted by Moskvin \etal (2002)
correspond to a polystyrene phantom.

For the 18~mm helmet (figure \ref{fig:201-18}), we found a good
agreement with the PENELOPE results of Moskvin \etal (2002) and with
those predicted by GP. Some differences with the calculation of
Xiaowei and Chunxiang (1999) appear at the ending edges of the plateau
of the maximum dose. For the 4~mm helmet (figure \ref{fig:201-04}),
the agreement with the GP predictions is rather good, while some
discrepancies are observed with the other calculations, mainly for the
$z$ profile (lower panel). Part of the disagreement with the PENELOPE
results of Moskvin \etal (2002) can be ascribed to the difference in
the material forming the phantom (polystyrene in the case of these
authors).

Due to lack of cylindrical symmetry shown by the source system of the
GK, an interesting point to address concerns the asymmetry shown by
the dose profiles. First, we have studied the asymmetry between $x$
and $y$ axes by means of $A_{xy}$ as given by equation
(\ref{eq:asymm}). In the upper panel of figure \ref{fig:asymm}, we
show the values obtained for the 18~mm treatment helmet. The shadow
regions indicate the $s$ values for which the corresponding dose
$D(s,0,0)$ is larger than 20\% (clearer) and 50\% (darker) of the
maximum dose, $[D(s,0,0)]_{\rm max}$. As we can see, the asymmetry
is below 15\% in absolute value for all $s$.  This percentage reduces
to around 5\% and 2\% in the two marked regions.

Also, we have determined for the same helmet, the asymmetries between
the negative and positive parts of the $x$ axis, $A_{x}$ (see equation
(\ref{eq:asymm-x})) and of the $y$ axis, $A_{y}$ (see equation
(\ref{eq:asymm-y})). Results are plotted in the lower panels of figure
\ref{fig:asymm}, where the shadow regions have the same meaning
mentioned above. Similar comments to those done for $A_{xy}$ can be
stated in both cases. We have checked that the situation is the same
for the remaining three helmets. The conclusion is that the loss of
cylindrical symmetry in the GK, provoked by the absence of some source
channels, has a rather slight effect on the dose profiles at the
isocenter. These profiles show up cylindrical symmetry in practice.

As a second test of our simplified model, we have performed new
simulations, in similar conditions to those of the previous
configuration, but plugging 51 and 99 sources, as indicated in the
schemes of figure \ref{fig:plugged}.

\begin{figure}
\begin{center}
\epsfig{figure=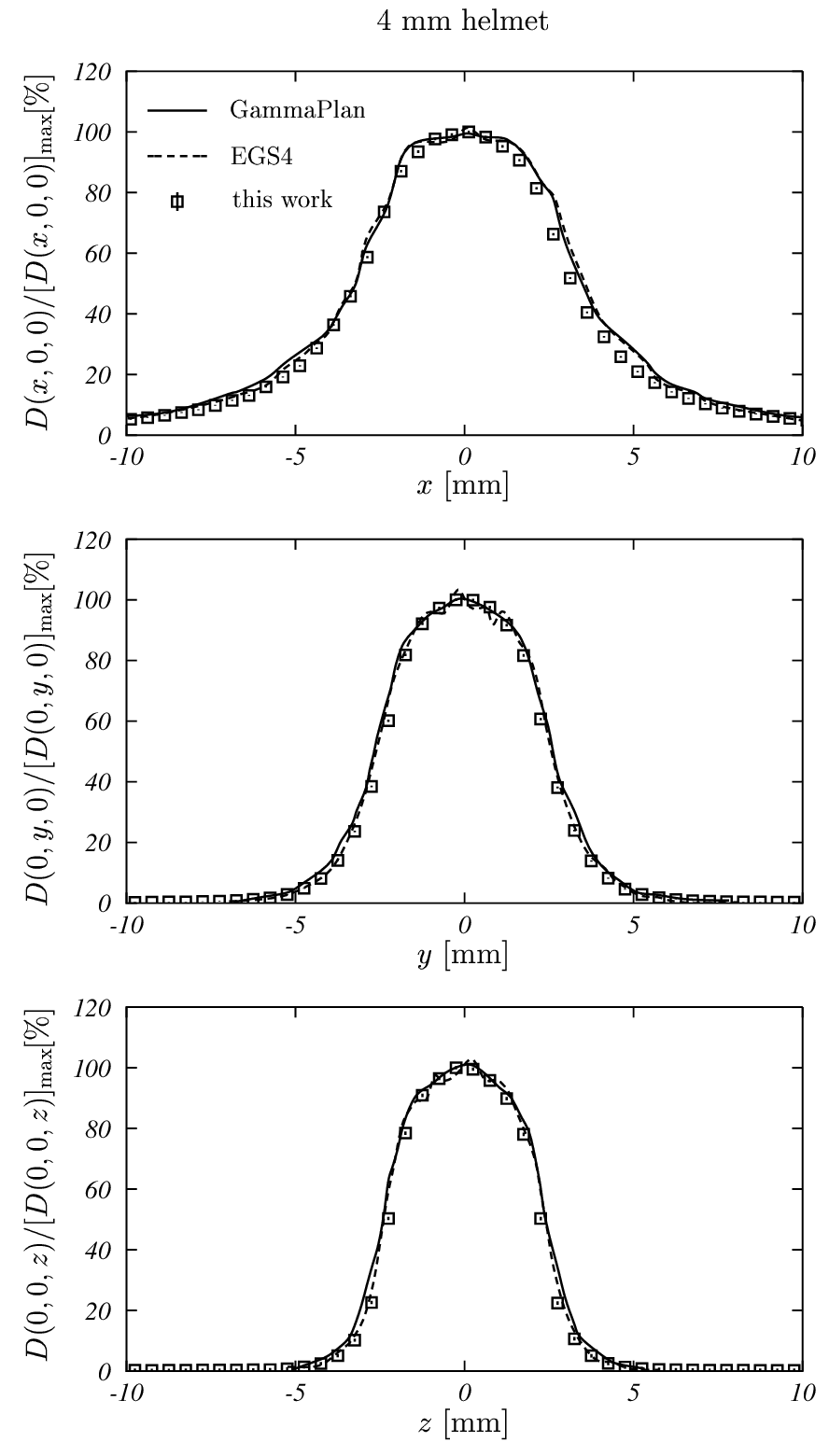,width=10cm}
\end{center}
\caption{Dose profiles at the isocenter, relative to their respective
maxima, in percentage, for the 4~mm helmet and for the configuration
with 150 unplugged sources. The results along $x$, $y$ and $z$ axes
are shown. Open squares are the results of our simulations. Dashed
curves have been obtained by Cheung \etal (1999a) with EGS4. Solid
curve corresponds to the predictions of GP quoted by Cheung \etal
(1999a). Both results have been obtained by scanning directly figures
3-8 of Cheung \etal (1999a). 
\label{fig:150-04}}
\end{figure}

\begin{figure}
\begin{center}
\epsfig{figure=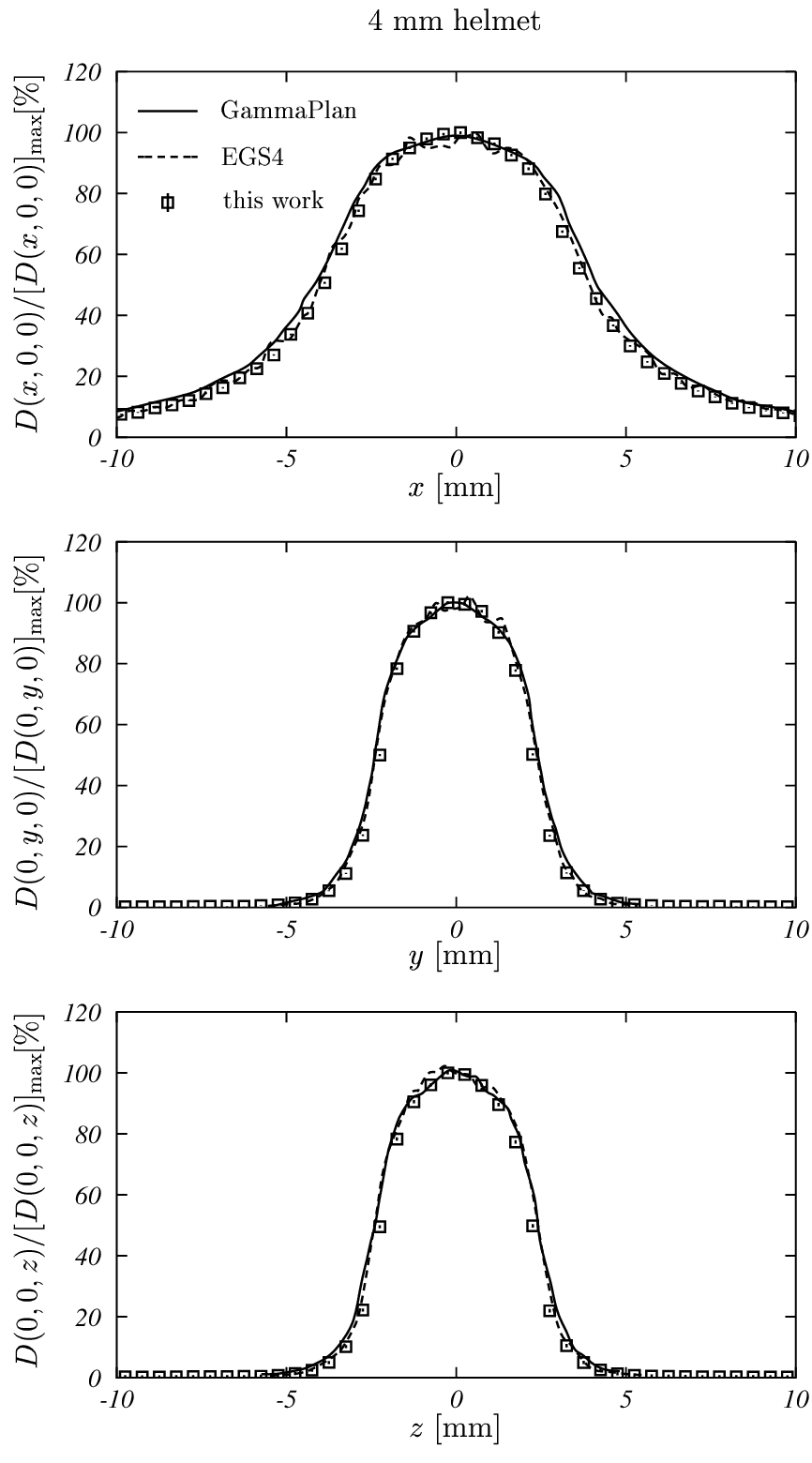,width=10cm}
\end{center}
\caption{Same as in figure \protect\ref{fig:150-04} but for the configuration
with 102 unplugged sources.
\label{fig:102-04}}
\end{figure}

In figures \ref{fig:150-04} and \ref{fig:102-04} we compare our
results for the 4~mm helmet (squares) with those of Cheung \etal
(1999a) obtained with EGS4 (dashed curve) and with the GP predictions
quoted by the same authors (solid curves).
The three profiles along $x$ (upper panels), $y$ (medium
panels) and $z$ (lower panels) axes are shown. As we can see, our
results show a very good agreement with those obtained with EGS4. On
the other hand, as in the case of the 201 source configuration, the
agreement with the GP predictions is rather good.

As for the case of a single source (Al-Dweri \etal (2004)), the
simplified model of the source channel produces dose profiles which
are in good agreement with other MC calculations and with
the GP predictions. This ensures the feasibility of the simplified
geometry model, which, in addition, permits a large reduction in the
computation time (larger than a factor 15) with respect to the
calculations with the full geometry. 

{\small
\Table{Comparison of the dose output factors quoted by different
authors and obtained with MC simulations or measured with various
techniques with those found in our calculations. The average value
quoted by Tsai \etal (2003) corresponds to a pool of measurements done
with silicon diode, diamond detector, radiographic film, radiochromic
film, and TLD cubes. The values of Moskvin \etal (2002) were obtained
with a spherical voxel of $R=$0.75~mm using a polystyrene phantom. The
values we have obtained correspond to the cubic voxel described in the
text (first row) and to spherical voxels with radii 0.5, 0.75 and
1~mm. In these three last cases a total of $2\cdot 10^8$ histories
have been followed.
\label{tab:output-factors}}
\br
 && 14~mm & 8~mm & 4~mm \\ 
\mr
Elekta (1992) &   & 0.984           & 0.956           & 0.800 \\
Elekta (1998) &   &                 &                 & 0.870 \\ 
\hline
Cheung \etal (1999b) & EGS4 &
   0.974$\pm$0.009 & 0.951$\pm$0.009 & 0.872$\pm$0.009 \\ \hline
                     & EGS4 &        & & 0.876$\pm$0.005 \\ 
                  & radiografic film &        & & 0.876$\pm$0.009 \\ 
Ma \etal (2000)   & radiochromic film &        & & 0.870$\pm$0.018 \\ 
                      & TLD &        & & 0.890$\pm$0.020 \\ 
                    & Diode &        & & 0.884$\pm$0.016 \\ \hline 
Moskvin \etal (2002) & PENELOPE  & &&&\\
 & $R=$0.75~mm & 
0.970$\pm$0.004 & 0.946$\pm$0.003 & 0.876$\pm$0.009 \\ \hline
Tsai \etal (2003) & average &
                 &                 & 0.868$\pm$0.014 \\ \hline
this work & PENELOPE & &&&\\
 & cubic voxel &
 0.982$\pm$0.007 & 0.967$\pm$0.007 & 0.876$\pm$0.006 \\
 & $R=$0.5~mm &
 0.99$\pm$0.03 & 0.95$\pm$0.03 & 0.86$\pm$0.03 \\
 & $R=$0.75~mm &
 0.99$\pm$0.02 & 0.96$\pm$0.02 & 0.86$\pm$0.01 \\
 & $R=$1~mm &
 0.978$\pm$0.009 & 0.950$\pm$0.008 & 0.846$\pm$0.006 \\
\br
\end{tabular} 
\end{indented}
\end{table}
}

Finally, we have calculated the dose output factors for the
configuration with 201 sources. First, we have performed the
calculations using the cubic voxel described in section 2.1. In table
\ref{tab:output-factors} we compare the results we have obtained (see
first row labeled ``this work'') with those found by other authors by
means of MC simulations or different measurement procedures. Our
results are in good agreement with the findings of the other authors
in case of the 4~mm helmet. The value we have obtained for the 8~mm
helmet agrees within the statistical uncertainties with the one quoted
by Cheung \etal (1999b), but it is noticeably larger than those of
Elekta (1992) and Moskvin \etal (2002). Finally, for the 14~mm helmet,
our result agrees with those of Cheung \etal (1999b) and the
manufacturer (Elekta 1992), but differs (at the 1$\sigma$ level) from
that of Moskvin \etal (2002).

Out of the discrepancies noted, the most significant are those found
with the calculations of Moskvin \etal (2002). These authors have used
the version 2000 of PENELOPE code, but the differences between this 
version and the 2001 we have used are not expected to produce such 
discrepancies. This is corroborated by the good agreement we have 
found when comparing the dose profiles discussed above. 

In order to clarify this disagreement, we have investigated if it is
due to differences in the scoring voxels chosen for the calculations.
Moskvin \etal (2002) considered a spherical voxel with radius
0.75~mm. Thus we have performed new simulations, following $2\cdot
10^8$ histories, and using the spherical voxels described in section
2.1. The results are shown in the last three rows of table
\ref{tab:output-factors}.  As we can see, there are not significant
variations (within the statistical uncertainties) when the radius of
the voxel is reduced. On the other hand, the results obtained for
these spherical voxels are into agreement with the values quoted by
Moskvin \etal (2002). This points out a dependence of the dose output
factors with the shape of the scoring voxel.

\section{Conclusions}

In this work we have investigated the dosimetry of the GK
by considering a simplified model for the single
source channels. Calculations have been done by using the Monte Carlo
code PENELOPE (v. 2001) for different configurations including 201, 
150 and 102 unplugged sources.  

The use of the simplified model produce results for the dose profiles
at the isocenter which are into agreement with previous
calculations done with other MC codes and with the predictions of the
GP. The absence of cylindrical geometry due to the lack of some source
channels in the GK does not show up in the calculated dose profiles.

Besides, we have determined the dose output factors corresponding to
the 14, 8 and 4~mm helmets. The results found show a good agreement
with those obtained with EGS4 and measured by means of different
procedures, mainly for the 4~mm helmet. The discrepancies observed
with previous results obtained also with PENELOPE are largely reduced
once one uses scoring voxels with the same shape. This voxel shape
dependence deserves a deeper investigation which we are carrying out
at present.

The results quoted here, together with those found for the single
source configuration (Al-Dweri \etal 2004), prove the suitability of
the simplified geometry proposed to perform dosimetry calculations for
the GK. The simplicity of this model and the level of accuracy which
can be obtained by using it opens the possibility to use MC tools for
planning purposes in the GK, mainly if we take into account the
reduction in the computational time (around a factor 15) with respect
to the full geometry simulations. As an additional gain, MC
simulations permit to take into account the presence of
inhomogeneities and interfaces in the target geometry, which are not
correctly treated by GP.

\ack{Authors wish to acknowledge M. Vilches for useful discussion and
G. Rey and A. Hamad for providing us with geometrical details of
the Leksell Gamma Knife$^{\circledR}$.  F.M.O. A.-D. acknowledges the
A.E.C.I. (Spain) and the University of Granada for funding his
research stay in Granada (Spain). This work has been supported in part
by the Junta de Andaluc\'{\i}a (FQM0220).}

\References

\item[] Al-Dweri F M O, Lallena A M and Vilches M 2004
A simplified model of the source channel of the Leksell
Gamma Knife$^{\circledR}$ tested with PENELOPE
{\it Phys. Med. Biol.}   {\bf 49} 2687-2703 

\item[] Cheung J Y C, Yu K N, Ho R T K and Yu C P 1999a
Monte Carlo calculations and GafChromic film measurements
for plugged collimator helmets of Leksell Gamma Knife
unit
{\it Med. Phys.} {\bf 26} 1252-6
                             
\item[] Cheung J Y C, Yu K N, Ho R T K and Yu C P 1999b
Monte Carlo calculated output factors of a Leksell Gamma Knife
unit
{\it Phys. Med. Biol.} {\bf 44} N247-9 

\item[] Cheung J Y C, Yu K N, Ho R T K and Yu C P 2000
Stereotactic dose planning system used in Leksell Gamma Knife
model-B: EGS4 Monte Carlo versus GafChromic films
MD-55
{\it Appl. Radiat. Isot.} {\bf 53} 427-30

\item[] Cheung J Y C, Yu K N, Yu C P and Ho R T K 1998
Monte Carlo calculation of single-beam dose profiles used in a gamma
knife treatment planning system
{\it Med. Phys.} {\bf 25} 1673-5

\item[] Cheung J Y C, Yu K N, Yu C P and Ho R T K 2001
Dose distributions at extreme irradiation depths of gamma knife
radiosurgery: EGS4 Monte Carlo
calculations 
{\it Appl. Radiat. Isot.} {\bf 54} 461-5

\item[] Elekta 1992 {\it Leksell Gamma Unit-User's Manual} (Stockholm:
Elekta Instruments AB)

\item[] Elekta 1996 {\it Leksell GammaPlan Instructions for Use for
Version 4.0-Target Series} (Geneva: Elekta)

\item[] Elekta 1998 {\it New 4-mm helmet output factor} 
(Stockholm: Elekta)

\item[]{Ma} 
Ma L, Li X A and Yu C X 2000 An efficient method of
measuring the 4 mm helmet output factor for the Gamma
Knife {\it Phys. Med. Biol.} {\bf 45} 729-733

\item[] Moskvin V, DesRosiers C, Papiez L, Timmerman R, Randall M and
DesRosiers P 2002 
Monte Carlo simulation of the Leksell Gamma Knife:
I. Source modelling and calculations in homogeneous media 
{\it Phys. Med. Biol.} {\bf 47} 1995-2011
 
\item[] Salvat F, Fern\'andez-Varea J M, Acosta E and Sempau J 2001
{\it PENELOPE, a code system for Monte Carlo simulation of
electron and photon transport} (Paris: NEA-OECD)

\item[] 
Tsai J-S, Rivard M J, Engler M J, Mignano J E, Wazer D E and 
Shucart W A 2003 Determination of the 4 mm Gamma Knife helmet relative 
output factor using a variety of detectors 
{\it Med. Phys.}  {\bf 30} 986-992

\item[] Xiaowei L and Chunxiang Z 1999 
Simulation of dose distribution irradiation by the Leksell Gamma Unit 
{\it Phys. Med. Biol.} {\bf 44} 441-5

\endrefs

\newpage

\end{document}